\documentclass[apj]{emulateapj}
\usepackage{amsmath}
\usepackage{graphicx}
\usepackage{amssymb}
\usepackage{times}

\begin{document}

\title{Nonlinear scattering of Fast Radio Bursts}

\author{Andrei Gruzinov}
\affiliation{Physics Dept., New York University,  726 Broadway, New York, NY 10003, USA}

\begin{abstract}
Nonlinear radio waves modulate the plasma, scatter on the modulations, and develop an intermittent power spectrum -- perhaps. Rudiments of theory, numerical simulations, and qualitative modeling of nonlinear scattering are presented. Nonlin ear radio waves in electron-positron plasma are shown to be unstable, the instability growth rate is calculated. The long-term outcome of the instability is studied numerically; an intermittent power spectrum is found. The frequency coherence scale of the intermittent power spectrum is interpreted as a measure of the nonlinear scattering screen.

\end{abstract}

\maketitle

\section{Introduction}
\label{sec:intro}

The great theoretical tools of radio astronomy DM and RM (dispersion and rotation measures) are based on full theoretical understanding of linear radio waves in plasma. Theory predicts the wave speed and the rate of Faraday rotation, giving frequency-dependent time delays and position angles at the observation point, resulting in the measurement of $\int dl~n=$ DM and $\int dl ~n~B_{\parallel}\propto$ RM. This is just two numbers, but still, we have learned something definite about the plasma.

In FRBs, the radio waves should be nonlinear close to the source \citep{Luan}, in the sense that 
\begin{equation}
\psi \equiv \frac{\lambda e E}{2\pi mc^2} \sim 1,
\end{equation}
where $\lambda$ is the wavelength, $e$, $m$ are the electron charge and mass, $E$ is the electric field. The nonlinearity parameter $\psi$ is proportional to the wave vector potential; $\psi$ is invariant under boosts along the wave; $\psi \sim 1$ means that an electron is accelerated to a relativistic velocity within a single wavelength. Nonlinearities $\psi \approx 0.1$ are indeed expected in the Simple Shmaser (SHock MASER) \citep{Gruz}. More sophisticated shmasers \citep{Waxm, GruW, Metz} must give similar $\psi$ inasmuch as they do describe FRBs. 

One wants to develop a theory for nonlinear radio waves and see if the nonlinearity produces observable features. Similar to DM and RM, if the nonlinear features are indeed detected, one wants to extract information about the plasma. 

The simplest possible theory of nonlinear radio waves is presented in this paper. The theory predicts an intermittent spectrum. The frequency coherence scale of the spectrum $\delta \nu$ is a measure of the nonlinear scattering screen: the screen thickness is 
\begin{equation}
\delta r\sim \frac{c}{\delta \nu}. 
\end{equation}

This proposition -- intermittent spectrum, with the frequency coherence scale given by the screen thickness -- is tentative, almost frivolous. We do not really prove it in what follows. The proposition is stated here mostly to stress the need for better understanding of nonlinear radio waves as regards their observational signatures. One needs to invent NM (nonlinearity measure) analogous to DM and RM. FRBs seem to be a good candidate for manifesting NM (incidentally, FRBs are also thought to be good for manifesting CM, conversion measure, a little-used  linear wave measure \citep{GruL}).

As far as this author can see, the most reliable way to better understand nonlinear radio waves is by massive PIC (Particle-In-Cell) numerical simulations. We expect that better understanding of nonlinear radio waves will considerably refine or even replace our proposition. But the intermittent spectrum may prove to be a robust prediction.

Intermittent spectra are normally attributable to scintillations, e.g. Fig.(5) of \cite{Hess} for FRB 121102. Scintillations probably render the current version of NM useless, unless the observed frequency dependence of the frequency coherence scale $\delta \nu _{\rm obs} (\nu)$  show clear deviations form the scintillation theory predictions. The chances of measuring the Nonlinear Measure would improve if we could predict $\delta \nu _{\rm NM}(\nu)$ theoretically, but so far we can't.

We consider the simplest plasma setup (\S\ref{sec:eq}) for the study of nonlinear radio waves (\S\ref{sec:per}). We show analytically that nonlinear radio waves are unstable and calculate the instability growth rate (\S\ref{sec:inst}). There is nothing principally new in \S\S\ref{sec:per}, \ref{sec:inst}. Exact solutions and instabilities of nonlinear radio waves have been known at least since the 1960s \citep{Zakh, Gorb, Max}. 

The question is what happens to the radio wave as the instability runs its course. We show numerically, by a PIC, that an intermittent spectrum of radio waves develops from a single initial mode  (\S\ref{sec:PIC}). Finally, we qualitatively study the resulting observed spectrum using a toy model (\S\ref{sec:rand}). 
 
In summary, nonlinear stages of FRBs or pulsar radio bursts are likely to leave observable imprints on the detected signal. We must try to identify the imprints and extract information about the plasma. Below is a crude attempt. 

\section{Nonlinear Radio Wave: Equations}
\label{sec:eq}

Here we write down the simplest possible system of equations describing nonlinear radio waves. We assume the fluid regime. We will generalize the equations to the kinetic regime later on, when it is needed.  

The simplest case is the planar linearly polarized wave in (initially) cold electron-positron plasma. Although we consider electron-positron plasma just for the sake of simplicity, the electron-positron plasma is, in fact, plausible; surely in pulsars, but also in FRBs, given the energetics. 

The wave is propagating along $x$, electric field $E$ along $y$, magnetic field $B$ along $z$. From $E$, electrons and positrons acquire equal in magnitude and oppositely directed $y$-velocities. Then from $B$ -- equal $x$-velocities. As the particles move in unison, we describe only one specie, say positrons
\begin{align}
\partial_t&u_x+v_x\partial_xu_x=Bv_y,\\
\partial_t&u_y+v_x\partial_xu_y=-Bv_x+E,\label{raw1}\\
\partial_t&n+\partial_x(nv_x)=0,\\
\partial_t&E=-\partial_xB-nv_y,\\
\partial_t&B=-\partial_xE.
\end{align}
Here $u$ is the 4-velocity, $v$ is the velocity, $n$ is the particle density (electrons plus positrons), and we use the ``natural units'':
\begin{equation}
c=\frac{e}{m}=4\pi e=1.
\end{equation}

Let $-\psi$ be the $y$-component of the vector potential. Then 
\begin{equation}
E=\partial_t\psi,~~~B=-\partial_x\psi,
\end{equation}
and Eq.(\ref{raw1}) is solved by (conservation of the $y$-component of generalized momentum)
\begin{equation}
u_y=\psi.
\end{equation}

Now, dropping the x-suffixes on $u_x$, $v_x$, we have the final system of equations describing nonlinear radio waves:
\begin{align}
\partial_t^2&\psi =\partial_x^2\psi -\frac{n}{\gamma}\psi,\label{bas1}\\
\partial_t&n+\partial_x(nv)=0,\label{bas2}\\
\partial_t&u+v\partial_xu=-\frac{1}{\gamma}\psi\partial_x\psi,\label{bas3}\\
\gamma^2&\equiv 1+u^2+\psi^2,~~~v\equiv \frac{u}{\gamma}.
\end{align}

\section{Nonlinear Radio Wave: Periodic Solutions}
\label{sec:per}
The best way to find the exact periodic nonlinear wave solution is to note that the basic system Eqs.(\ref{bas1}-\ref{bas3}) is 1+1 Lorentz-covariant:
\begin{align}
(\partial_a&\partial^a+n_0)\psi=0,\\
\partial_a&(n_0u^a)=0,\\
u^b&\partial_bu_a=\psi\partial_a\psi,
\end{align}
where $\psi$ and the proper density $n_0\equiv \frac{n}{\gamma}$ are Lorentz scalars, and $u^a\equiv(\gamma,u)$ is a Lorentz vector. 

From Eqs.(\ref{bas1}-\ref{bas3}), we get the standard linear wave dispersion law
\begin{equation}
\omega^2=k^2+n_0\equiv k^2+\omega_p^2.
\end{equation}
Since $\omega>k$, the phase speed is greater than $1$, and one can boost to the photon rest frame, where $k=0$ and $\psi$ depends only on time. 

Now we drop the linearity assumption, and consider the general $x$-independent solution of Eqs.(\ref{bas1}-\ref{bas3}):
\begin{align}
\ddot{\psi}&=-\frac{n\psi}{\gamma},\label{nosc1}\\
\gamma^2&\equiv 1+u^2+\psi^2, ~~~n={\rm const},~~~u={\rm const}.\label{nosc2}
\end{align}

For arbitrary constants $n$, $u$, Eqs.(\ref{nosc1}, \ref{nosc2}) give nonlinear oscillations of arbitrary amplitude. The solution can be arbitrarily boosted, giving the general periodic nonlinear wave.

\section{Nonlinear Radio Wave: Instability}
\label{sec:inst}
Here we show that nonlinear periodic radio waves in electron-positron plasma are unstable and calculate the instability growth rate. A rigorous procedure for calculating the instabilities is described in \S\ref{sec:rig}. The rigorous calculation can only be done numerically and is not very illuminating. In \S\ref{sec:appr} we do an approximate calculation which explains the nature of the instability and gives a simple analytic expression for the growth rate.

\subsection{Instability: Exact Calculation}\label{sec:rig}
In the wave rest frame, the unperturbed background is the time-dependent nonlinear wave Eqs.(\ref{nosc1}, \ref{nosc2}). This is an exact solution of Eqs.(\ref{bas1}-\ref{bas3}). Now consider perturbations of the exact solution 
\begin{equation}
\delta \psi,~\delta n, ~\delta u~\propto e^{ikx}.
\end{equation}
To linear order, Eqs.(\ref{bas1}-\ref{bas3}) give
\begin{align}
\ddot{\delta \psi}&=-\left( k^2+\frac{1+u^2}{\gamma^3}n\right) \delta\psi-\frac{\psi}{\gamma}\delta n+\frac{nu\psi}{\gamma^3}\delta u,\\
\dot{\delta n}&=ik\frac{nu\psi}{\gamma^3}\delta\psi-ik\frac{u}{\gamma}\delta n-ikn\frac{1+\psi^2}{\gamma^3}\delta u,\\
\dot{\delta u}&=-ik\frac{\psi}{\gamma}\delta\psi-ik\frac{u}{\gamma}\delta u.
\end{align}

One solves this system of equations simultaneously with  Eq.(\ref{nosc1}) for one period of the nonlinear oscillator Eq.(\ref{nosc1}). This should be repeated four times, with different initial conditions for the perturbations 
\begin{equation}
\delta \psi,~\dot{\delta \psi},~\delta n, ~\delta u.
\end{equation}
Relating the perturbations before and after one oscillation period gives the transfer matrix, four-by-four. The instability growth rate is given by the real part of the logarithm of the eigenvalue of the transfer matrix divided by the oscillator period. 

We did it numerically and found an instability for large enough $k$. The numerical results agree with the analytic results given in the next subsection.

\subsection{Instability: Approximate Calculation}\label{sec:appr}
The nature of the instability is best understood, and the growth rates are most easily calculated, if one limits the accuracy to cubic order in $\psi$. Noting that $u$ and $n-n_0$ are quadratic in $\psi$, one readily derives the simplified system of equations:
\begin{align}
\partial_t^2&\psi =\partial_x^2\psi -n_0(1+\phi-\frac{1}{2}\psi^2)\psi,\label{tr1}\\
\partial_t^2&\phi =\frac{1}{2}\partial_x^2\psi^2 \label{tr2},
\end{align}
where 
\begin{equation}
n\equiv (1+\phi)n_0.
\end{equation}

The linear waves of this system are the electromagnetic waves,
\begin{equation}
\omega^2=k^2+n_0,~~~\phi=0,
\end{equation}
and the density ``waves'',
\begin{equation}
\omega^2=0,~~~\psi=0.
\end{equation}

Now consider a high-$k$ density wave perturbation of a low-frequency, $\omega \ll k$, electromagnetic wave. To leading order in large $k$, to second order in the unperturbed electromagnetic wave $\psi$, to linear order in perturbations $\delta\psi$, $\delta\phi$:
\begin{align}
\partial_x^2\delta\psi&\approx n_0\psi \delta\phi,\\
\partial_t^2\delta\phi&\approx \psi\partial_x^2\delta\psi,
\end{align}
giving
\begin{equation}
\partial_t^2\delta\phi \approx n_0\psi^2\delta \phi,
\end{equation}
with the growth rate
\begin{equation}
\omega _I\approx <\psi^2>^{1/2}\omega_p.
\end{equation}

The nature of the instability must be as follows. The background electromagnetic wave drives currents along $y$. Like currents attract, the density modulations grow -- essentially the Weibel instability. 

The growth rate of the nonlinear wave instability should be compared to the induced Compton scattering rate (\cite{Weym, Belo} and references therein; written in the form of a growth rate):
\begin{equation}
\omega _{I~{\rm ICS}}\sim <\psi^2>\frac{\omega_p^2}{\omega}.
\end{equation}

For the Simple Shmaser, $\psi \sim 0.1$, $\omega\sim\omega_p$, the nonlinear wave instability growth rate is somewhat larger than the induced scattering rate. 

\section{Kinetic description of Nonlinear Radio Waves}
\label{sec:PIC}

Now, knowing that nonlinear waves are unstable, an obvious thing to do is to study the full nonlinear system Eqs.(\ref{bas1}-\ref{bas3}) or its cubic-order truncation Eqs.(\ref{tr1},\ref{tr2}) numerically and see how the spectrum evolves with time. We did it, but only to discover, as we should have anticipated, that this approach does not work. As the instability develops, the fluid description breaks down -- the particle trajectories cross. The nonlinear evolution must be studied in kinetics. It's a pity, but things are as they are.

\begin{figure}
\plotone{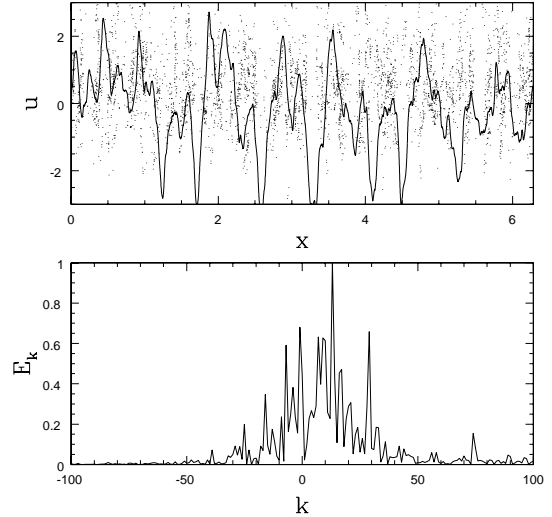}
\caption{Upper panel: $x$ and $y$ components of the 4-velocity, $u$ and $\psi$, vs $x$. For $u_x$ only every 10th particle is shown. For $u_y$ we just plot $\psi$ as a function of $x$. Lower panel: Fourier transform of electromagnetic wave energy. Negative $k$ correspond to left-going waves.}
\label{kinf}
\end{figure}

The fluid system Eqs.(\ref{bas1}-\ref{bas3}) can be readily generalized to the full kinetic regime, when particles are described by the distribution function $f=f(t,x,u)$:
\begin{align}
\partial_tf&+\frac{1}{\gamma}u\partial_xf-\frac{1}{\gamma}\psi\partial_x\psi\partial_uf=0,\label{kin1}\\
\partial_t^2\psi&=\partial_x^2\psi-\left( \int du\frac{1}{\gamma}f\right) \psi,~~~\gamma^2\equiv 1+u^2+\psi^2.\label{kin2}
\end{align}

Numerical simulations, by a PIC, of the system Eqs.(\ref{kin1},\ref{kin2}) were performed. In a $2\pi$-periodic box, we initiate a nonlinear wave with the following initial conditions: at $t=0$,
\begin{equation}
n=30,~~~\psi=2,~~~\dot{\psi}=0,~~~u=1.
\end{equation}
What we see at $t=1.6$ is shown in Fig.(\ref{kinf}). Note the highly intermittent spectrum of electromagnetic waves, which has developed from the initially single mode $k=0$. 

Besides the intermittent spectrum, another important, and very much unanticipated, result of our PIC simulations is the longevity of nonlinear electromagnetic waves. The wave does go unstable with the growth rate $\sim \omega _I\approx <\psi^2>^{1/2}\omega_p$, but we find that for $\psi \sim 1$, $\omega\sim\omega_p$,  even after $t\gtrsim 1000\omega_p^{-1}$, the electromagnetic wave survives, keeping most of the energy, and only broadening the spectrum to $\delta\omega\sim \omega_p$.

Unfortunately, this author cannot run a realistic PIC simulation of nonlinear scattering, even in 1+1 dimensions. For a 1 ms long FRB at 1 GHz, a million wavelengths thick scattering screen is needed. What's shown in Fig.(\ref{kinf}) corresponds to just a few tens of waves. We offer a qualitative model of nonlinear scattering in \S\ref{sec:rand}.

\section{Scattering in a random medium}
\label{sec:rand}
A nonlinear beam of electromagnetic waves modulates the plasma of the scattering screen. Both the plasma density and velocity are modulated, and the modulations are time-dependent. We will consider a qualitative model, where the time-dependence and the velocity modulations are ignored, and we simplify to 1+1 dimensions:
\begin{equation}
\partial_t^2\psi=\partial_x^2\psi-n\psi, \label{rscat}
\end{equation}
where $n=n(x)$ is a given random density field.
With the mean plasma frequency as a unit of measurement,
\begin{equation}
\omega_p\equiv<n>^{1/2}=1,
\end{equation}
we have three dimensionless parameters characterizing the scattering problem Eq.(\ref{rscat}): 
\begin{enumerate}
\item $L$, the thickness of the screen;
\item $l_c$, the coherence length of the density modulations;
\item $\frac{\delta n}{n}$, the modulation amplitude.
\end{enumerate}

\begin{figure}
\plotone{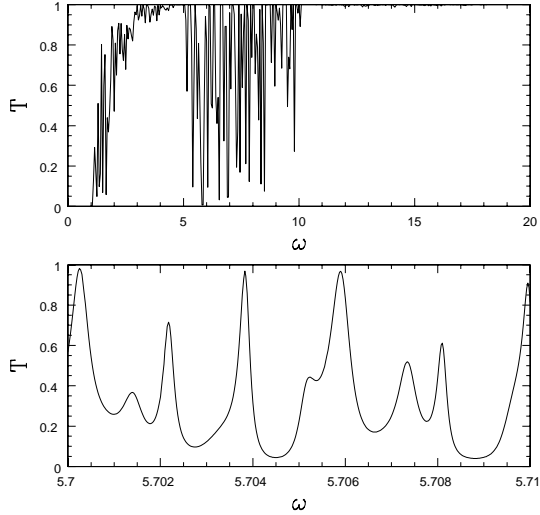}
\caption{Upper panel: transmission coefficient vs frequency. Lower panel: Blow-up of the upper panel.}
\label{scatf}
\end{figure}

Fig.(\ref{scatf}) shows the transmission coefficient of the screen with
\begin{equation}
L\approx 3000,~~~l_c\approx 0.3,~~~\frac{\delta n}{n}\approx 0.5.
\end{equation}

Note the highly intermittent transmission at $\omega \lesssim 10$, with the frequency coherence scale $\delta \omega \sim 0.001$. This is superficially similar to Fig.(5) of \cite{Hess}, with $\sim 100$ kHz frequency coherence scale at $\sim 1$ GHz. But scintillations are probably the dominant, if not the only, source of the observed intermittency \citep{Hess}. 

What allows us to claim that the scattering in a random medium problem Eq.(\ref{rscat}) can be a reasonable proxy for nonlinear scattering is the relationship
\begin{equation}
\omega l_c \sim 1,
\end{equation}
satisfied for $\omega \sim 10$, where the intermittent spectrum is seen. This is to be expected if the waves themselves are responsible for the density modulations. 

The cutoff at $\omega =\omega_p=1$ is due to the fact that the minimal frequency of the wave capable of propagating inside the screen is $\omega_p$. The origin of the frequency coherence scale $\delta \omega \sim 0.001$ seen at $\omega \sim 10$ needs explaining.

To this end, consider perturbative scattering, with $\frac{\delta n}{n}\ll 1$. In this regime, the reflection coefficient can be calculated in the Born approximation 
\begin{equation}
R_\omega =|r_\omega|^2,~~~r_\omega=\frac{1}{2\omega}\int dx~n(x)e^{-2i\omega x},
\end{equation}
and the transmission coefficient is
\begin{equation}
T_\omega=1-R_\omega.
\end{equation}

The frequency coherence scale of the transmission coefficient is given by the wavenumber coherence scale of the density Fourier transform $n_k$. We will model the density field in the screen by
\begin{equation}
n(x)\propto V(x)\phi(x),
\end{equation}
where
\begin{equation}
V(x)\propto e^{-\frac{x^2}{2L^2}},
\end{equation}
is the smooth envelope of thickness $\sim L$, and $\phi$ is a Gaussian random field with coherence length $\sim l_c$: 
\begin{equation}
<\phi(x)\phi(y)>\propto e^{-\frac{(x-y)^2}{2l_c^2}}.
\end{equation}
The correlator of the density Fourier transform is now easily computed. For $l_c\ll L$:
\begin{equation}
<n_kn_{k'}^*>\propto \exp \left( -\frac{1}{2}l_c^2k^2-\frac{1}{4}L^2(k-k')^2 \right).
\end{equation}
This gives the frequency coherence scale of the transmission coefficient 
\begin{equation}
\delta \omega \sim \frac{1}{L},
\end{equation}
in agreement with Fig.(\ref{scatf}). 

\section{Discussion}
\label{sec:disc}

Tentatively, with high uncertainty, we propose that: (i) Nonlinear scattering of radio waves gives rise to intermittent power spectra. (ii) The frequency coherence scale $\delta \nu$ of the power spectrum gives the thickness $\delta r$ of the nonlinear scattering screen $\delta r\sim \frac{c}{\delta \nu}$. (iii) As nonlinear scattering is most pronounced near the source, the thickness of the nonlinear scattering screen is expected to be comparable to the size of the emission region. One should see structures within the burst on the time scale $\tau \sim \frac{1}{\delta\nu}$. (iv) Massive PIC simulations of nonlinear radio waves are probably needed to confirm and refine -- or replace -- our proposition.

\vskip 0.5cm

\acknowledgements 

I thank Jason Hessels and Harish Vedantham for many private lessons in basic radio astronomy, and Andrei Beloborodov, Peter Goldreich, and Yuri Levin for useful discussions of nonlinear radio waves.

\vskip 1cm

\bibliographystyle{hapj}

\end{document}